%% file: main.tex
\documentclass[conference]{IEEEtran}
\IEEEoverridecommandlockouts
\PassOptionsToPackage{table, x11names, dvipsnames, svgnames}{xcolor}
\usepackage{cite}
\usepackage{bbm}
\usepackage{amsmath,amssymb,amsfonts}
\usepackage{mathtools}
\usepackage{graphicx}
\usepackage{booktabs}
\usepackage{textcomp}
\usepackage{xcolor}
\usepackage[english]{babel}
\usepackage[acronym,nonumberlist]{glossaries}
\usepackage{glossaries-prefix}
\usepackage{pgfplots}
\pgfplotsset{compat=1.17}
\usepgfplotslibrary{fillbetween}
\usepackage{amssymb}
\usepackage{import}
\usepackage{tabularx}
\usepackage{caption}
\usepackage{algorithmic}
\usepackage{algorithm}
\usepackage{bm}
\usepackage{subfig}
\usepackage[edges]{forest}
\forestset{fold/.style={folder, grow'=0, s sep=0.1pt}, 
    arr/.style={edge=-latex}}
\usepackage{tikz}
\usepackage{siunitx}
\usepackage{multirow}
\usepackage{hyperref}
\usepackage{braket} 
\hypersetup{%
linktocpage=true, 
colorlinks=false,
pdfborder={0 0 0}, 
breaklinks=true, pdfpagemode=UseNone, pageanchor=true, pdfpagemode=UseOutlines,%
plainpages=false, bookmarksnumbered, bookmarksopen=true, bookmarksopenlevel=1,%
hypertexnames=true, pdfhighlight=/O,
pdftitle={CVRP and circuit cutting},%
pdfauthor={Christian Ufrecht},%
pdfsubject={QuaST},%
pdfkeywords={},%
pdfcreator={pdfLaTeX},%
pdfproducer={LaTeX with hyperref}%
}
\usepackage[inline]{enumitem}

\usepackage[capitalise]{cleveref}

\usepackage{xcolor}

\usepackage{calc}

\begin{document}



\title{Unitary Synthesis of Clifford+T Circuits with Reinforcement Learning
\thanks{

* These authors contributed equally (name order randomized). \\
The research is part of the Munich Quantum Valley (MQV), which is supported by the Bavarian state government with funds from the Hightech Agenda Bayern Plus.\\
Email address for correspondence:\\
abhishek.yogendra.dubey@iis.fraunhofer.de}
}

\author{\IEEEauthorblockN{Sebastian Rietsch\IEEEauthorrefmark{1}, Abhishek Y. Dubey\IEEEauthorrefmark{1}, Christian Ufrecht, Maniraman Periyasamy, \\
Axel Plinge, Christopher Mutschler, and Daniel D. Scherer}
\IEEEauthorblockA{
Fraunhofer Institute for Integrated Circuits IIS,\\
Nordostpark 84, 90411 Nürnberg, Germany}
}

\maketitle

\input{glossary.tex}

\begin{abstract}
    This paper presents a deep reinforcement learning approach for synthesizing unitaries into quantum circuits. Unitary synthesis aims to identify a quantum circuit that represents a given unitary while minimizing circuit depth, total gate count, a specific gate count, or a combination of these factors. While past research has focused predominantly on continuous gate sets, synthesizing unitaries from the parameter-free Clifford+T gate set remains a challenge. Although the time complexity of this task will inevitably remain exponential in the number of qubits for general unitaries, reducing the runtime for simple problem instances still poses a significant challenge. In this study, we apply the tree-search method Gumbel AlphaZero to solve the problem for a subset of exactly synthesizable Clifford+T unitaries. Our method effectively synthesizes circuits for up to five qubits generated from randomized circuits with up to $60$ gates, outperforming existing tools like QuantumCircuitOpt and MIN-T-SYNTH in terms of synthesis time for larger qubit counts. Furthermore, it surpasses Synthetiq in successfully synthesizing random, exactly synthesizable unitaries.
    These results establish a strong baseline for future unitary synthesis algorithms.
\end{abstract}


\section{Introduction}
\label{sec:intro}
\input{Introduction}

\section{Fundamentals}
\label{sec:fundamentals}
\input{Fundamentals}

\section{Related work}
\label{sec:related-work}
\input{RelatedWork}

\section{Gumbel AlphaZero}
\label{sec:method}
\input{GumbelAlphaZero}

\section{Our RL framework}
\label{sec:environment}
\input{RLUnitaryCompil}

\section{Experiments}
\label{sec: exps}
\input{Experiments}


\section{Outlook}
\label{sec:outlook}
In our presentation of our \gls{rl} framework, we focused on the exact synthesis of unitaries using the Clifford+T gate set. With some further improvements, the scope of our agent's application can be expanded. Firstly, our approach does not preserve the MCTS tree following the execution of actions, meaning the search tree is rebuilt from scratch at every iteration. We aim to address this limitation in our future work. A further enhancement could involve adopting a more advanced neural network design, like a transformer network, which could consider the circuit's configuration in addition to the target unitary, potentially elevating the agent's efficacy. An intriguing inquiry is whether a more effective method exists for representing the unitary as input to the neural network.
We would also like to alleviate a limitation of our benchmarking results by comparing other RL based methods for synthesizing unitaries such as Ref.~\cite{lu2023qas}.

Transitioning to approximate synthesis theoretically entails dropping the constraint that $C_{HS}$ be precisely zero for two unitaries, which will be the next obvious step. Another consideration is to incorporate unitaries that utilize ancilla qubits, which could potentially reduce the depth of the synthesized circuit.

\section{Conclusion}
\label{sec:conclusion}
Traditional heuristic approaches to unitary synthesis rely on manually designed rules to reduce search space effectively. On the other hand, automated methods utilizing \gls{rl} endure the initial cost of extensive training periods but can independently discover these synthesis strategies. Our empirical findings indicate that the inference times did not markedly increase with the number of qubits, marking a notable improvement over traditional heuristic techniques. Therefore, our proposed method establishes a new benchmark for unitary synthesis algorithms that employ reinforcement learning. 

\section*{Acknowledgement}
We are very grateful to P.~ Mukhopadhyay for helpful discussions and for sharing the code with us, which was used to generate the 
results for the MIN-T-SYNTH method shown in  \cref{tab:SpecificUnitaries}.  We would also like to thank V.~Gheorghiu for sharing the 
Github repository with us for the approximate circuit synthesis in Clifford+T gates. We thank N.~Quetschlich for
interesting discussions related to quantum compilation using reinforcement learning. Finally, we would like to thank Q.~Göttl for comments on the manuscript.

\bibliographystyle{IEEEtran}
\bibliography{references}

\appendix
\begin{appendices}
\section*{Quantum circuit optimization}
\label{sec: appendix1}
\input{ApxQCOpt}

\end{appendices}

\end{document}

%% file: glossary.tex
\newacronym{rl}{RL}{reinforcement learning}
\newacronym{ml}{ML}{machine learning}
\newacronym{dnn}{DNN}{deep neural network}
\newacronym{qc}{QC}{quantum computing}
\newacronym{vqc}{VQC}{variational quantum circuit}
\newacronym{mdp}{MDP}{Markov Decision Process}
\newacronym{mcts}{MCTS}{Monte Carlo tree search}
\newacronym{dqn}{DQN}{Deep Q-Networks}
\newacronym{ppo}{PPO}{Proximal Policy Optimization}
\newacronym{nisq}{NISQ}{noisy intermediate-scale quantum}
\newacronym{hst}{HST}{Hilbert-Schmidt Test}

%% file: Introduction.tex
The advent of quantum computing foreshadows a paradigm shift in our computational capabilities, sparking significant efforts towards scalable implementations of quantum algorithms on actual quantum hardware.
Like classical computers, executing a quantum algorithm on a device requires converting a high-level algorithmic representation into low-level, hardware-dependent operations that can be executed on the device.
The umbrella term for this task is quantum circuit compilation, a broad set of different constituents, encompassing optimization, qubit mapping and routing, gate scheduling, error correction, and classical control integration, among others~\cite{maronese2022quantum}.

One of the sub-tasks of quantum compilation for a problem-specific unitary matrix is to identify a quantum circuit comprised of gates from a pre-defined gate set whose corresponding unitary matrix closely approximates or precisely matches it, known as \emph{approximate} and \emph{exact} \emph{unitary synthesis} respectively.
Our focus in this paper is on synthesizing quantum circuits from unitary matrices, an important sub-routine in any quantum development kit, that can generate efficient quantum circuits for targeted hardware~\cite{davis2020towards, younis2021qfast, smith2023leap}.

In the fault-tolerant setting~\cite{shor1996fault}, the Clifford+T gate set plays a crucial role due to its efficiency in realizing fault-tolerant quantum computation. The T-gate is essential for enabling universal quantum computing, but it is considered costly because of the necessity of magic state distillation~\cite{campbell2017n}, which introduces significant resource overhead and increases circuit depth~\cite{eastin2009restrictions}. Similarly, in \glspl{nisq} devices~\cite{preskill2018quantum}, two-qubit entangling gates are typically more error-prone and have longer execution times compared to single-qubit gates~\cite{masanes2002time}. To reduce overall error rates and improve the reliability of synthesized circuits, synthesis algorithms must take into account the specific characteristics of certain gates, such as the T-gate in fault-tolerant circuits and two-qubit gates in \gls{nisq} devices. By minimizing the use of these costly gates, the synthesis algorithms can generate circuits that are more resilient to errors and better suited for their respective computing paradigms.

Although exact and provably optimal synthesis methods for Clifford+T circuits exist~\cite{amy2013meet,mosca2021polynomial}, their practicality is limited due to runtime complexities that grow exponentially with the number of qubits and the number of T-gates. This limitation highlights the need to explore heuristic approaches to make unitary synthesis more applicable to real-world problems where the circuit representation is unknown, such as problem-specific quantum oracles. Recent developments in \gls{rl} have shown that it is a highly successful framework for learning such heuristics through directed trial and error. Inspired by initial contributions towards \gls{rl}-based synthesizers~\cite{moro2021cp}, we model unitary synthesis as a sequential decision-making problem in which an \gls{rl} agent appends one gate at a time to the quantum circuit. Although model-free \gls{rl} has proven successful on simple tasks~\cite{moro2021cp,he2021njp}, its application remains limited to circuits with a small number of qubits and gate sets.

In recent years, model-based \gls{rl} utilizing \gls{mcts}~\cite{coulom2007cg, kocsis2006mle2}, such as AlphaGo~\cite{silver2016n} and AlphaZero~\cite{silver2018s}, have shown great promise in solving complex problems with large search spaces~\cite{fawzi2022n, mankowitz2023n, trinh2024n}. Recently, the Gumbel AlphaZero algorithm~\cite{danihelka2022iclr} has proven effective in solving complex problems while requiring fewer computational resources compared to its predecessors. 

To address the limitations of past unitary synthesis methods, we propose a novel method that leverages the strengths of Gumbel AlphaZero and incorporates domain-specific knowledge to enable efficient and scalable unitary synthesis. In our study, we address the exact synthesis of Clifford+T unitaries, even though, in theory, our method is not limited to one specific gate set or an exact synthesis regime. 
We show that our method scales up to 5-qubit circuits and can, on average, synthesize unitaries into circuits comprised of fewer T-gates than those present in the circuits used to generate the target unitary. 
Furthermore, when the agent is tasked with synthesizing structured unitaries such as the Toffoli and Peres gates, we show that it performs optimally in terms of the number of T-gates for almost all examples.

In contrast to other approaches that are trained for one specific unitary~\cite{he2021njp, daimon2023a}, our agent can synthesize a wide range of input unitaries without the need for retraining, highlighting the effectiveness and generalizability of our Gumbel AlphaZero-based approach.
To the best of the authors' knowledge, we are the first to scale \gls{rl}-based unitary synthesis past the one- or two-qubit setting with a generalizing agent.

The paper is organized as follows: \Cref{sec:fundamentals} introduces unitary synthesis and \gls{rl}. \Cref{sec:related-work} reviews previous work in unitary synthesis using classical heuristics and \gls{rl} methods. \Cref{sec:environment} describes our \gls{rl} environment formulation. \Cref{sec:method} presents our Gumbel AlphaZero-based algorithm. \Cref{sec: exps} details the experimental setup and discusses the results, and \Cref{sec:outlook} explores future improvements. The Appendix provides a brief overview of quantum circuit optimisation.

%% file: Fundamentals.tex
\subsection{Quantum Unitary Synthesis}
\label{sec:fundamentals_compilation}

Unitary synthesis is the process of devising an efficient quantum circuit that implements a given unitary matrix, which can then be processed by a quantum compiler for execution on a real device. Formally, for a gate set $\mathcal{G}$, a circuit $C$ is said to be over $\mathcal{G}$ if it consists solely of gates from $\mathcal{G}$. We denote the set of all such circuits as $\langle \mathcal{G} \rangle$. Given a unitary $U \in \mathcal{U}(2^n)$, a gate set $\mathcal{G}$, and a precision $\epsilon$, the goal of unitary synthesis is to find a circuit $C \in \langle \mathcal{G} \rangle$ whose unitary representation $V_C$ satisfies $d(U, V_C) \leq \epsilon$, where $d: \mathcal{U}(2^n) \times \mathcal{U}(2^n) \rightarrow \mathbb{R}$ is a distance metric. Setting $\epsilon$ to zero defines the problem of \emph{exact synthesis}.

\textit{a) Distance metrics: } The Hilbert-Schmidt distance, average gate fidelity, and trace distance are among the most commonly used metrics for comparing unitaries. A detailed discussion of these and other metrics used in quantum information theory is available in Ref.~\cite{wilde2013quantum}.

In our approach, we utilize the Hilbert-Schmidt inner product (HS), defined as $\langle U, V_t \rangle = \mathrm{Tr}(V_t^{\dagger}U)$, to measure the similarity between a target unitary $U$ and another unitary $V_t$, which represents the synthesized circuit at time step $t$. The cost function for optimization in a $2^n$-dimensional space, corresponding to $n$ qubits, is given by:
\begin{equation}
    \label{eq:hilbert-schmidt}
    C_{HS} = 1 - \frac{1}{4^n}|\text{Tr}(V_t^{\dagger}U)|^2
\end{equation}
Note that this cost function is invariant to the global phase, meaning unitaries are considered equivalent under this metric, even if they differ by a global phase.

\textit{b) Clifford+T: } 
The Clifford+T gate set, essential for many fault-tolerant quantum computing protocols, includes the Clifford gates, which map Pauli strings to Pauli strings up to a phase of $\{ \pm 1, \pm i \}$, and the T-gate. However, as already mentioned, implementing error-correcting codes in fault-tolerant quantum computers poses a challenge due to the T-gate, which necessitates minimizing its occurrences in a circuit~\cite{eastin2009restrictions}. 
In cases where an arbitrary unitary cannot be exactly synthesized with Clifford+T, a sequence of Clifford+T gates can still be found to approximate the unitary to any desired level of accuracy~\cite{dawson2005solovay}.

To quantify the optimality of T-gate usage, we adopt the \emph{T-count} metric, $\mathcal{T}(U)$. It is defined as the minimum number of T-gates required to implement a unitary $U$ up to a global phase for unitaries that can be exactly synthesized from the Clifford+T gate set~\cite{gosset2013algorithm, giles2013exact, gheorghiu2022t}.

\subsection{Reinforcement Learning}
Recent advancements in \gls{rl} have led to its application in quantum computing for tasks such as quantum compilation, circuit optimization, and routing~\cite{moro2021cp, furrutter2023quantum, weiden2023improving, liang2023unleashing, preti2023hybrid, paler2023machine, fosel2021quantum, zen2024quantum, olle2023simultaneous, ruiz2024quantum}.

\gls{rl} is a learning paradigm, where an agent learns to optimize a reward signal through iteratively interacting with an environment along some time horizon. The agent can perform actions that may be discrete, continuous, or a combination of both. 
The problem is formally modelled as a \gls{mdp}, characterized by a set of states $\mathcal{S}$, a set of actions $\mathcal{A}$, a reward function $r(s,a,s')$, a discount factor $\gamma$, and an environment dynamics function $p(s'| s, a)$, describing the probability of transitioning to state $s'$ given the current state $s$ and action $a$. 

The primary goal is to devise an optimal policy $\pi^*(a|s)$, a probabilistic function mapping states to action probabilities, that maximizes the expected sum of discounted rewards, called the \emph{expected return}, for every possible environment state. 
Given the reward trajectory of an episode, the return at time step $t$ is defined as $G_t = \sum_{k=t+1}^T \gamma^{k-t-1}r_k$, where $r_k$ is the reward at time $k$, and $T$ is the episode's length. Given a policy $\pi$ and a general state $s$, we denote the expected return of $s$ under $\pi$ as $v_{\pi}(s)$, also known as the \emph{value} function. The expected return of taking an action $a$ in a state $s$ under $\pi$, denoted $q_{\pi}(s, a)$, is referred to as the \emph{state-action} or \emph{Q-value} function. In the following, we work with estimators of Q-value function and denote it as ${Q}(s,a)$. For more details on \gls{rl}, refer to Ref.~\cite{andrew2018reinforcement}.

%% file: RelatedWork.tex
\subsection{Classical Unitary Synthesis}
\label{sec: classical}

Pioneering theoretical contributions to the exact synthesis of single-qubit unitaries using the Clifford+T gate set were made by Kliuchnikov et al.~\cite{kliuchnikov2012fast}. Their work centered on determining the possibility of exact unitary synthesis and devising an efficient sequence of gates for implementation. Giles and Selinger further extended these concepts to general $n$-qubit unitaries, introducing an exact synthesis algorithm~\cite{giles2013exact}, which was later refined by Kliuchnikov~\cite{kliuchnikov2013synthesis}. Although these methods provided foundational insights, they did not prioritize T-gate reduction or gate count minimization, and both exhibited exponential run-time complexities.
Selinger first proposed a highly efficient algorithm tailored for approximating arbitrary single-qubit unitaries with the Clifford+T gate set~\cite{selinger2012efficient}, marking a significant stride towards practical unitary synthesis.

The first deterministic method aimed at finding the T-count of a unitary which is exactly synthesizable using Clifford+T gate set was introduced by Gosset et al.~\cite{gosset2013algorithm}. This method leveraged the channel representation of unitaries and a decision algorithm for determining the T-count of a unitary $U$, denoted as $\mathcal{T}(U) \leq m$, for a specified integer $m$. The complexity of this algorithm was characterized as $\mathcal{O}(2^{nm} \text{poly}(m, 2^n))$, with $n$ representing the number of qubits and $m$ the T-count.

The first heuristic algorithm for synthesizing a T-count optimal unitary was developed by Mosca and Mukhopadhyay~\cite{mosca2021polynomial}. Their algorithm, \emph{MIN-T-SYNTH}, prunes the search space by exploiting the structure of the channel representation of the unitary, achieving a conjectured polynomial complexity relative to the T-count. Despite its advancements, the algorithm's exponential complexity relative to the number of qubits remains a challenge.

\emph{QuantumCircuitOpt}~\cite{QCOpt_SC2021} offers a practical approach to unitary synthesis by converting the circuit design task into specialized mixed-integer programs (MIP)~\cite{bixby2004mixed}, solvable using MIP solvers such as Gurobi~\cite{gurobi}. This framework is versatile, accommodating various gate sets beyond Clifford+T, including parameterized gates and higher-level gates like controlled-V and Toffoli. However, exact synthesis with minimal T-gates is practically feasible only for low-depth circuits with a small number of qubits.

Finally, Synthetiq~\cite{paradis2024synthetiq} is a recently published tool used to synthesize circuits over arbitrary finite gate sets using Simulated annealing (SA). The approach is to iteratively modify a randomly built circuit, until a given criterion is met. The tool is versatile in doing exact synthesis using Clifford+T and other custom finite gate sets, approximate synthesis, synthesis for partially specified operators and also use composite gates such as the Toffoli gate in the gate set for synthesis.

\subsection{Reinforcement Learning-Based Unitary Synthesis}

Given the large design space of quantum circuits, the automation of unitary synthesis via reinforcement learning has been the subject of various studies. An initial theoretical investigation into the feasibility of modelling quantum state preparation and gate synthesis as a Markov Decision Process (MDP) was conducted in Ref.~\cite{alam2019}, showing that optimal gate sequences could be discovered using this framework for both noisy and noiseless single-qubit scenarios.

\gls{rl} has since been applied to a range of quantum circuit synthesis problems, particularly those involving parameterized quantum circuits. For example, double Q-learning was utilized for the variational compilation of specific 2 and 3-qubit unitaries in Ref.~\cite{he2021njp}, including the controlled-H, quantum Fourier transform, Toffoli, and W gates. Additionally, Monte Carlo tree search methods (MCTS) have been employed for hybrid optimization and problem-specific quantum circuit design, as seen in Refs.~\cite{wang2023itqe, yao2022pmsml}. Other research has focused on hardware-specific quantum circuit design using reinforcement learning, as detailed in Refs.~\cite{baum2021pq, kuo2021, li2023qip, preti2023}.

A recent study has further demonstrated the use of \gls{rl} in the synthesis of quantum circuits, specifically targeting state preparation~\cite{kolle2024reinforcement}. This research explored the relationship between the depth of synthesized circuits, the depth of circuits used for initialization during training, and the number of qubits. The study showed promising results in synthesizing simple 2-qubit product and Bell states.

For approximate synthesis, Moro et al. pioneered the concept of synthesizing quantum circuits from fixed unitaries as a sequential decision-making process~\cite{moro2021cp}. Employing DQN~\cite{mnih2013playing} and PPO~\cite{schulman2017proximal}, they successfully synthesized general single-qubit Haar-unitaries under the HRC efficient universal set~\cite{harrow2002efficient} and a rotational gate set. Chen et al. extended this approach to two-qubit unitaries using a combination of DQN and A* search, albeit with a search complexity that grows exponentially with the number of gates required~\cite{chen2022}. Further stride towards approximately synthesizing unitaries for more than 2 qubits using RL was done in Ref.~\cite{lu2023qas}, although this was not the main aim of the work. 

In this paper, we explore the application of reinforcement learning to unitary synthesis, aiming to extend its scalability to circuits with more qubits and greater depth.

%% file: GumbelAlphaZero.tex
Although the AlphaZero algorithm~\cite{silver2018s} has demonstrated significant success, it can encounter difficulties when the \gls{mcts} simulation budget is limited, which may lead to under-representation of certain actions and an inadequate enhancement of the policy network that directs the search. 
To overcome this, the Gumbel AlphaZero framework~\cite{danihelka2022iclr} was developed, which excels in scenarios where the number of possible discrete actions $n_{a}$ in a certain state greatly exceeds the simulation budget $n_{sim}$.

Gumbel AlphaZero employs the Gumbel-Top-$k$ trick at the root node to sample a subset of $k$ actions from a total of $n_a$ possible actions without replacement. 
This method utilizes a vector $g \in \mathbb{R}^{n_a}$ of independent Gumbel-distributed random variables and a policy network $\pi_{\theta}$, where $\pi_{\theta}(s_0) = \text{logits} \in \mathbb{R}^{n_a}$ for the root state $s_0$. 
The Gumbel-Top-$k$ trick selects the actions that maximize the expression $g + \text{logits}$, denoted as $\mathrm{argtop}\;(g + \text{logits}, k)$, where $\mathrm{argtop}$ returns the indices of the $k$ largest values of the vector.
Actions are then chosen from the sub-sampled set according to
\begin{equation} \label{eq:gaz-as}
	A = \underset{a}{\mathrm{argmax}} \; (g(a) + \text{logits}(a) + \sigma (Q(s_0, a)),
\end{equation}
where $Q(s_0, a) \in \mathbb{R}^{n_a}$ represents the empirical state-action values, $\sigma$ is any arbitrary monotonically increasing function, and $A$ is the action selected. 
The authors prove that selecting $A_{n_{sim}+1}$ as per \cref{eq:gaz-as} guarantees policy improvement, given the estimate of the state-action values for the root state is accurate. They instantiate $\sigma$ as
\begin{equation}
\sigma(Q(s_0, a)) = (c_{\text{visit}} + \underset{a'}{\mathrm{max}} N(s_0, a'))c_{\text{scale}} Q(s_0, a)
\end{equation}
inspired by the MPO policy update~\cite{abdolmaleki2018}, where $N(s_0, a)$ denotes the visitation count for action $a$ at root state $s$, and $c_{\text{visit}}$ and $c_{\text{scale}}$ are hyper-parameters. 
Additionally, sequential halving~\cite{karnin2013p3icicml-v2} is applied to efficiently allocate the simulation budget across the sampled actions, offering a bandit algorithm that focuses on minimizing simple regret, in contrast to the cumulative regret minimization in AlphaZero.
To reduce variance in the Q-value estimates, action selection at non-root nodes is deterministic, diverging from the stochastic selection used in the standard AlphaZero formulation.

%% file: RLUnitaryCompil.tex
\begin{figure}
    \begin{center}
        \def\svgwidth{\columnwidth}
        \centerline{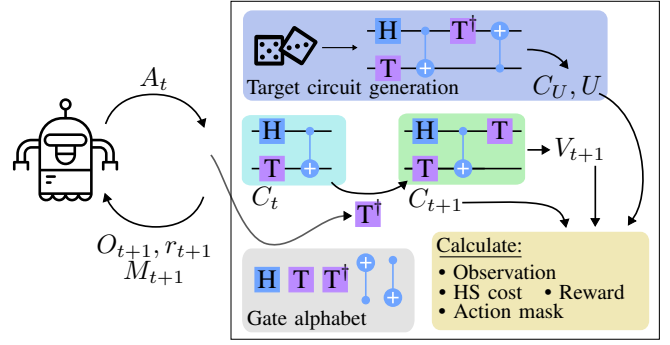}
        \caption{\textbf{Schematics of the \gls{rl} environment.} An episode begins by sampling a random circuit $C_U$ from the Clifford+T gate set to establish the target unitary $U$ for the episode. The state of the episode is initialized with an empty circuit. At each step $t$: (1) The circuit $C_t$ is augmented by appending the gate corresponding to action $A_t$, yielding a new circuit $C_{t+1}$ with its associated unitary $V_{t+1}$; (2) The target unitary $U$ and the new unitary $V_{t+1}$ are compared using the metric defined in \cref{eq:hilbert-schmidt}; (3) The reward $r_{t+1}$ is calculated based on the comparison from step (2); (4) Any operations that would lead to gate cancellations or redundancies in $C_{t+1}$ are marked as invalid for the subsequent step; (5) The agent receives the next observation $O_{t+1}$, the reward $r_{t+1}$, and the action mask $M_{t+1}$.}
        \label{fig:rl_env}
    \end{center}
\end{figure}

Since we focus on exact synthesis, we assume from the outset that the target unitary can be exactly synthesized using the Clifford+T gate set. Our corresponding \gls{rl} framework is depicted in~\cref{fig:rl_env} and, in its essence, aligns with the model proposed by Moro et al.~\cite{moro2021cp}.

\textit{a) MDP Definition: }
The agent is tasked with the iterative construction of a quantum circuit that embodies a unitary $V$, which should coincide with a predefined target unitary $U$. In the following, we denote by $V_t$ the unitary representation of the current circuit $C_t$ at time step $t$.
Actions within our framework correspond to the selection and addition of a gate from the gate set to the circuit at the earliest vacant moment for the required qubit(s).
Each potential placement of a gate constitutes a distinct action. 
For instance, with single-qubit gates in an $n$-qubit circuit, there are $n$ possible actions, one for each qubit.
For the synthesis to be considered exact, the cost function (as specified in~\cref{eq:hilbert-schmidt}) must be exactly zero, only allowing the circuit to induce a global phase difference.

The agent incurs a penalty of $-1$ for every gate added to the circuit and receives a reward of $0$ upon exact synthesis of the target unitary.
This incentivizes the agent towards shorter episodes and thus minimizes the total number of gates placed\footnote{An alternative is to penalize the agent for every $T/T^{\dagger}$ gate added to the circuit. In our experiments, this gave worse total gate count with no improvement in the number of T-gates.}. 
An episode concludes once the target unitary has been successfully synthesized or when a preset maximum number of steps is exceeded.
For the observation of the agent, we use $UV_t^{\dagger}$ instead of, for example, the concatenation of $U$ and $V_t$.
This is expected to stabilize training by i) normalizing the goal state observation to the identity matrix independent of the target $U$ and ii) alleviating the curse of dimensionality.
Intuitively, $UV_t^{\dagger}$ can be interpreted as the unitary residual toward the target unitary $U$, meaning the unitary of the yet unbuilt part of the final circuit.

\textit{b) Generation of Target Unitary: }
To generate target unitaries during training, we initiate each episode by sequentially building a random quantum circuit from the Clifford+T gate set and then calculating the unitary matrix $U$ of the resultant target circuit $C_U$.
The synthesis task's complexity is presumed to increase with the increment in the number of gates comprising the target circuit.
To facilitate learning, we confine the gate count of sampled circuits to a specified interval with a feasible upper limit.

\begin{figure}[ht] 
	\vskip 0.2in
	\begin{center}
		\centerline{\includegraphics[width=\columnwidth]{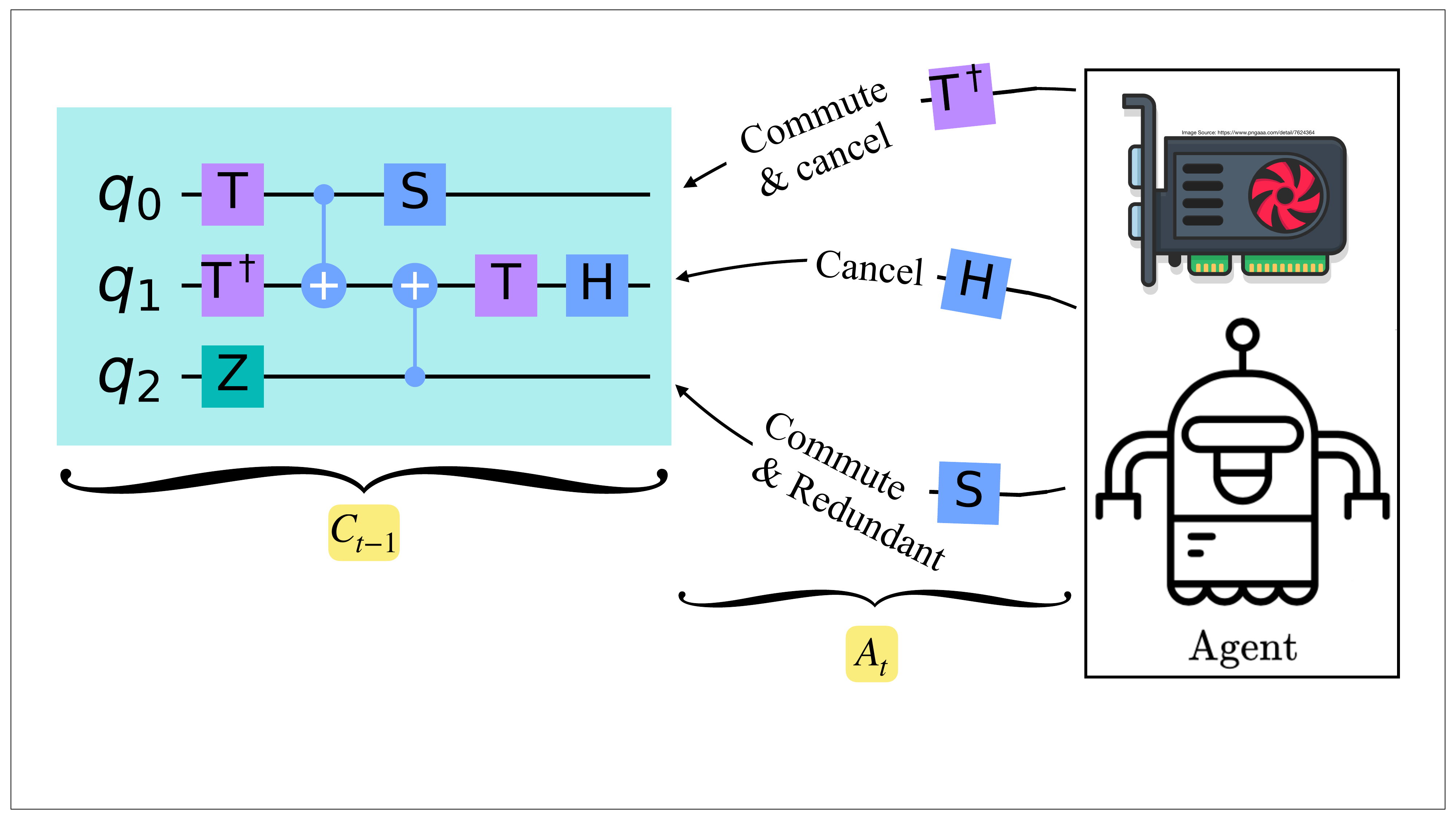}}
		\caption{Example of action constraints: The agent has to check the circuit $C_{t-1}$, for various cancellations and redundancies before appending the next gate. A $T^{\dagger}$ gate cannot be place on the first qubit since it commutes all the way to the first moment and cancels the $T$ gate. An $H$ gate cannot be placed on the second qubit since it cancels the already present $H$ gate. This is done until no more cancellation/commutations are possible.}
		\label{fig:action_mask_sch}
	\end{center}
	\vskip -0.2in
\end{figure}

\textit{c) Constraining the Search Space: }
Our approach employs tree search to tackle the challenge of synthesizing unitary operations. In particular, we implement Gumbel AlphaZero to conduct a parallel search, aiming to identify the optimal circuit within the set of potential solutions for a given unitary operation.
To alleviate the search complexity, we enforce state-dependent constraints on the set of actions that the agent can select at each time step~\cite{nagarajan20212isiwqcsq}.
More specifically, given that our method is based on a tree search strategy, our goal is to avoid cycles in the state evolution that can arise from the effects of gate cancellations.
Consider, for instance, the scenario where two successive gates negating each other are applied, such as an H-gate followed by another H-gate on the same qubit.
This sequence of operations would return the system to its prior state, rendering the exploration of such branches in the tree futile.
Similarly, our goal is to avoid gate sequences that could be implemented by a single gate from the original gate set. 
For example, applying a T-gate followed by another T-gate on the same qubit is equivalent to applying a single S-gate.

Given that many gates from the Clifford+T set commute, we assess these conditions for all gates within the circuit that commute to the earliest available circuit moment.
More formally, for two unitary gates $A, B \in \mathcal{G}$, with $\mathcal{G}$ denoting the Clifford+T gate set, we say that $A$ and $B$ commute iff $AB = BA$.
Assuming gate $A$ is part of the circuit $C_{t-1}$ and commutes to the earliest vacant circuit moment, we invalidate the action of appending a gate $B$ to the circuit at time step $t$ when either of the following conditions is satisfied:
\begin{enumerate}
    \item \textbf{Cancellation}: $BA = I$, where $I$ is the identity matrix.
    \item \textbf{Redundancy}: $BA = C$, where $C \in \mathcal{G}$ is a gate within the gate set.
\end{enumerate}
 
We provide an example of these constraints in~\cref{fig:action_mask_sch}. While not comprehensive, our action masking approach is anticipated to significantly reduce the search complexity.
To provide challenging targets for the agent during training, we employ the same masking strategy during the sampling of a target unitary $U$.

%% file: rl-env_v2.pdf_tex
\begingroup%
  \makeatletter%
  \providecommand\color[2][]{%
    \errmessage{(Inkscape) Color is used for the text in Inkscape, but the package 'color.sty' is not loaded}%
    \renewcommand\color[2][]{}%
  }%
  \providecommand\transparent[1]{%
    \errmessage{(Inkscape) Transparency is used (non-zero) for the text in Inkscape, but the package 'transparent.sty' is not loaded}%
    \renewcommand\transparent[1]{}%
  }%
  \providecommand\rotatebox[2]{#2}%
  \newcommand*\fsize{\dimexpr\f@size pt\relax}%
  \newcommand*\lineheight[1]{\fontsize{\fsize}{#1\fsize}\selectfont}%
  \ifx\svgwidth\undefined%
    \setlength{\unitlength}{294.29269625bp}%
    \ifx\svgscale\undefined%
      \relax%
    \else%
      \setlength{\unitlength}{\unitlength * \real{\svgscale}}%
    \fi%
  \else%
    \setlength{\unitlength}{\svgwidth}%
  \fi%
  \global\let\svgwidth\undefined%
  \global\let\svgscale\undefined%
  \makeatother%
  \begin{picture}(1,0.51158031)%
    \lineheight{1}%
    \setlength\tabcolsep{0pt}%
    \put(0,0){\includegraphics[width=\unitlength,page=1]{rl-env_v2.pdf}}%
    \put(0.63274089,0.13644962){\color[rgb]{0,0,0}\makebox(0,0)[lt]{\lineheight{1.25}\smash{\begin{tabular}[t]{l}{\footnotesize Calculate:}\end{tabular}}}}%
    \put(0,0){\includegraphics[width=\unitlength,page=2]{rl-env_v2.pdf}}%
    \put(0.34914046,0.37261641){\color[rgb]{0,0,0}\makebox(0,0)[lt]{\lineheight{1.25}\smash{\begin{tabular}[t]{l}{\footnotesize Target circuit generation}\end{tabular}}}}%
    \put(0,0){\includegraphics[width=\unitlength,page=3]{rl-env_v2.pdf}}%
    \put(0.35652581,0.20632425){\color[rgb]{0,0,0}\makebox(0,0)[lt]{\lineheight{1.25}\smash{\begin{tabular}[t]{l}$C_{t}$\end{tabular}}}}%
    \put(0.77597608,0.37305573){\color[rgb]{0,0,0}\makebox(0,0)[lt]{\lineheight{1.25}\smash{\begin{tabular}[t]{l}$C_U, U$\end{tabular}}}}%
    \put(0.80838244,0.28029328){\color[rgb]{0,0,0}\makebox(0,0)[lt]{\lineheight{1.25}\smash{\begin{tabular}[t]{l}$V_{t+1}$\end{tabular}}}}%
    \put(0,0){\includegraphics[width=\unitlength,page=4]{rl-env_v2.pdf}}%
    \put(0.37979129,0.08276754){\color[rgb]{0,0,0}\makebox(0,0)[t]{\lineheight{1.25}\smash{\begin{tabular}[t]{c}H\end{tabular}}}}%
    \put(0,0){\includegraphics[width=\unitlength,page=5]{rl-env_v2.pdf}}%
    \put(0.4308601,0.08276754){\color[rgb]{0,0,0}\makebox(0,0)[t]{\lineheight{1.25}\smash{\begin{tabular}[t]{c}T\end{tabular}}}}%
    \put(0,0){\includegraphics[width=\unitlength,page=6]{rl-env_v2.pdf}}%
    \put(0.38734018,0.24859982){\color[rgb]{0,0,0}\makebox(0,0)[t]{\lineheight{1.25}\smash{\begin{tabular}[t]{c}T\end{tabular}}}}%
    \put(0,0){\includegraphics[width=\unitlength,page=7]{rl-env_v2.pdf}}%
    \put(0.48543438,0.08243523){\color[rgb]{0,0,0}\makebox(0,0)[t]{\lineheight{1.25}\smash{\begin{tabular}[t]{c}$\text{T}^{\dagger}$\end{tabular}}}}%
    \put(0.21112628,0.38409874){\color[rgb]{0,0,0}\makebox(0,0)[t]{\lineheight{1.25}\smash{\begin{tabular}[t]{c}$A_t$\end{tabular}}}}%
    \put(0.21178217,0.13458684){\color[rgb]{0,0,0}\makebox(0,0)[t]{\lineheight{1.25}\smash{\begin{tabular}[t]{c}$O_{t+1}, r_{t+1}$\end{tabular}}}}%
    \put(0.21111851,0.09582519){\color[rgb]{0,0,0}\makebox(0,0)[t]{\lineheight{1.25}\smash{\begin{tabular}[t]{c}$M_{t+1}$\end{tabular}}}}%
    \put(0,0){\includegraphics[width=\unitlength,page=8]{rl-env_v2.pdf}}%
    \put(0.38709132,0.30454013){\color[rgb]{0,0,0}\makebox(0,0)[t]{\lineheight{1.25}\smash{\begin{tabular}[t]{c}H\end{tabular}}}}%
    \put(0,0){\includegraphics[width=\unitlength,page=9]{rl-env_v2.pdf}}%
    \put(0.6158121,0.30454013){\color[rgb]{0,0,0}\makebox(0,0)[t]{\lineheight{1.25}\smash{\begin{tabular}[t]{c}H\end{tabular}}}}%
    \put(0,0){\includegraphics[width=\unitlength,page=10]{rl-env_v2.pdf}}%
    \put(0.56060815,0.39836244){\color[rgb]{0,0,0}\makebox(0,0)[t]{\lineheight{1.25}\smash{\begin{tabular}[t]{c}T\end{tabular}}}}%
    \put(0,0){\includegraphics[width=\unitlength,page=11]{rl-env_v2.pdf}}%
    \put(0.616061,0.24859982){\color[rgb]{0,0,0}\makebox(0,0)[t]{\lineheight{1.25}\smash{\begin{tabular}[t]{c}T\end{tabular}}}}%
    \put(0,0){\includegraphics[width=\unitlength,page=12]{rl-env_v2.pdf}}%
    \put(0.72766505,0.30454013){\color[rgb]{0,0,0}\makebox(0,0)[t]{\lineheight{1.25}\smash{\begin{tabular}[t]{c}T\end{tabular}}}}%
    \put(0.59015396,0.20420023){\color[rgb]{0,0,0}\makebox(0,0)[lt]{\lineheight{1.25}\smash{\begin{tabular}[t]{l}$C_{t+1}$\end{tabular}}}}%
    \put(0.34964222,0.02800767){\color[rgb]{0,0,0}\makebox(0,0)[lt]{\lineheight{1.25}\smash{\begin{tabular}[t]{l}{\footnotesize Gate alphabet}\end{tabular}}}}%
    \put(0,0){\includegraphics[width=\unitlength,page=13]{rl-env_v2.pdf}}%
    \put(0.53437777,0.18011063){\color[rgb]{0,0,0}\makebox(0,0)[t]{\lineheight{1.25}\smash{\begin{tabular}[t]{c}$\text{T}^{\dagger}$\end{tabular}}}}%
    \put(0,0){\includegraphics[width=\unitlength,page=14]{rl-env_v2.pdf}}%
    \put(0.67595371,0.45396552){\color[rgb]{0,0,0}\makebox(0,0)[t]{\lineheight{1.25}\smash{\begin{tabular}[t]{c}$\text{T}^{\dagger}$\end{tabular}}}}%
    \put(0,0){\includegraphics[width=\unitlength,page=15]{rl-env_v2.pdf}}%
    \put(0.56035932,0.45430274){\color[rgb]{0,0,0}\makebox(0,0)[t]{\lineheight{1.25}\smash{\begin{tabular}[t]{c}H\end{tabular}}}}%
    \put(0.63591942,0.06645005){\color[rgb]{0,0,0}\makebox(0,0)[lt]{\lineheight{1.25}\smash{\begin{tabular}[t]{l}{\footnotesize • HS cost}\end{tabular}}}}%
    \put(0.79386241,0.06650631){\color[rgb]{0,0,0}\makebox(0,0)[lt]{\lineheight{1.25}\smash{\begin{tabular}[t]{l}{\footnotesize • Reward}\end{tabular}}}}%
    \put(0.63540493,0.0953354){\color[rgb]{0,0,0}\makebox(0,0)[lt]{\lineheight{1.25}\smash{\begin{tabular}[t]{l}{\footnotesize • Observation}\end{tabular}}}}%
    \put(0.63585629,0.03767721){\color[rgb]{0,0,0}\makebox(0,0)[lt]{\lineheight{1.25}\smash{\begin{tabular}[t]{l}{\footnotesize • Action mask}\end{tabular}}}}%
    \put(0,0){\includegraphics[width=\unitlength,page=16]{rl-env_v2.pdf}}%
  \end{picture}%
\endgroup%

%% file: Experiments.tex
In this section, we present the experimental results of our \gls{rl} agent for synthesizing unitaries and describe our setup in \cref{sec:experiments_setup}.
We evaluate the agent on two types of unitaries: random unitaries in \cref{sec: exps2} and structured unitaries in \cref{sec: exps1}.
We benchmark our agent against state-of-the-art synthesis algorithms MIN-T-SYNTH~\cite{mosca2021polynomial}, QuantumCircuitOpt~\cite{nagarajan20212isiwqcsq} and Synthetiq~\cite{paradis2024synthetiq}.

\subsection{Implementation and Setup}
\label{sec:experiments_setup}
Our environment code was fully implemented in JAX~\cite{deepmind2020jax}, leveraging its JIT (just-in-time) compilation to address the inherent complexities of the training and inference process and enable seamless agent parallelization on a GPU. To ensure correctness, the calculation of the unitary using our framework for a given circuit was cross-validated with the comprehensive quantum computing software Qiskit~\cite{Qiskit}.
For Gumbel AlphaZero, we utilize an implementation from the DeepMind mctx library~\cite{danihelka2022iclr} for all of our experiments, which is also implemented in JAX.

We employ simple feed-forward neural networks for the actor and critic model.
Each network is composed of five layers, with 1024 hidden units per layer, and incorporates layer normalization~\cite{ba2016layer} for regularization purposes~\cite{bjorck2022}.
To improve stability during the training of the critic network, we normalized the critic outputs into the range of $[-1, 1]$.
To process the complex-valued observations $UV_t^{\dagger}$, we simply split the real and imaginary parts into separate entries and flatten the resulting matrix.

\textit{a) Training: } 
Each qubit model (2-5 qubits) is trained for different durations as shown in~\cref{tab:training_table}.
The training utilizes a replay buffer with the capacity to store 100,000 transitions. Here, we distribute the data collection across 256 concurrent environment workers and compute the training step gradients using a batch size of 1024.
As a result, each transition sample has an update-to-data ratio of four.
We sample the gate count of target circuits uniformly from the range $[3, \beta]$ where $\beta$ depends on the number of qubits as shown in~\cref{tab:training_table}.
We reduce $\beta$ for 5 qubits to alleviate the exponential growth in the size of the observation and action spaces.
The gate set of our experiments is $\mathcal{G} = \{ H, T, T^{\dagger}, S, S^{\dagger}, Z, CX \}$.
We highlight that, \emph{although the initial training takes a considerable amount of time, a trained model can be subsequently used multiple times for synthesis in a compilation pipeline for any exactly synthesizable unitary}.

\begin{table}
\renewcommand{\arraystretch}{1.3}
\caption{Table of training times and range of sampling gates for the model we use to synthesize unitaries}
\label{tab:training_table}
\centering
\begin{tabular}{c||c||c||c}
\hline
\bfseries Qubits & \bfseries \begin{tabular}[c]{@{}l@{}}Number of time steps\\ (in millions)\end{tabular} & \bfseries \begin{tabular}[c]{@{}l@{}}Training\\ time (days)\end{tabular} & \bfseries \begin{tabular}[c]{@{}l@{}}Range of\\ sampling gates\end{tabular} \\
\hline\hline
2 & 2 & 14 & [3,60]\\
\hline
3 & 1.6 & 10.33 & [3,60]\\
\hline
4 & 1 & 7.5 & [3,60]\\
\hline
5 & 1 & 6.66 & [3,40]\\
\hline
\end{tabular}
\end{table}



\textit{b) Parallel search using Gumbel AlphaZero: }
Given a Gumbel AlphaZero model that has been adequately trained, meaning weights $\theta, 
\mu$ of the policy network $\pi_{\theta}$ and the value network $\upsilon_{\mu}$ are available, we implement a parallelized search strategy.
In contrast to sampling actions greedily with Gumbel samples set to zero leveraging a single agent, we follow the same strategy of Gumbel-based action sampling of training with multiple stochastic agents deployed in parallel. From the candidate solutions, we select the one with the highest score, which is determined by a scoring function that evaluates the quality of the solution (either the total gate count or the number of T-gates).

This procedure takes advantage of the variability stemming from the Gumbel distribution to investigate a wider array of possible solutions, increasing the diversity of the search trajectories and improving the likelihood of discovering superior circuits.
Further, parallelizing the search not only reduces the chance of sub-optimal solutions but also expedites the search by making efficient use of available computational resources. 
Most importantly, it allows the natural incorporation of two-level optimization goals within the search process: minimizing the overall gate count, encouraged by a $-1$ reward penalty for every gate added, and minimizing the number of T-gates by scanning the collection of identified solution circuits in a subsequent step.
The specifics of our algorithm are presented in \cref{algo:mcts-eval}, 
with the internal data structure of our implementation below it.

\begin{algorithm}[t]
    \caption{Parallel search using Gumbel AlphaZero}
    \label{algo:mcts-eval}
    \begin{algorithmic}[1]
        \REQUIRE $n_a$: number of actions; $n_{sim}$: number of simulations; $k$: number of actions to sample; $b$: number of runs
        \REQUIRE $\pi_{\theta}$: policy network; $\upsilon_{\mu}$: value network
		\REQUIRE $score(x)$: Solution scoring function
		\REQUIRE $s_0$: Initial state of the environment to solve
        \STATE Replicate initial state $s_0$ $b$ times: $\vec{s} \gets [s_0^0, s_0^1, ..., s_0^{b-1}]$
        \WHILE{not all finished for every non-terminal run $i$}
            \STATE $s_0 \gets \vec{s}_i$
            \STATE Compute $\text{logits}$ using $\pi_{\theta}(s_0)$
            \STATE Sample Gumbel variables $(g \in \mathbb{R}^{n_{a}}) \sim \text{Gumbel}(0)$
            \STATE Find $k$ actions using the Gumbel-top-$k$ trick \newline $\mathcal{A}_{\text{topk}} = \mathrm{argtop} \; (g + \text{logits}, k)$
            \STATE Use sequential halving with $n_{sim}$ MCTS simulations, root action selection $g(a) + \text{logits}(a) + \sigma(Q(s_0, a))$, deterministic non-root action selection and $\upsilon_{\mu}$ for leaf node evaluation
			\STATE $A_{n_{sim} + 1} = \underset{a \in \text{Remaining}}{\mathrm{argmax}}(g(a) + \text{logits}(a) + \sigma(Q(s_0, a)))$
			\STATE Step the environment with action $A_{n_{sim} + 1}$ and update to successor state $s'$: $\vec{s}_i \gets s'$
        \ENDWHILE
        \STATE Identify runs with valid solutions and their indices $I_{valid}$
		\RETURN $\underset{i \in I_{valid}}{\mathrm{argmax}} \; score(\vec{s}_i)$
    \end{algorithmic}
\end{algorithm}

\begin{algorithm}[t]
\caption*{Environment Data Structures}
\label{alg:data-structures}
\begin{algorithmic}[1]
    \STATE \textbf{Data Structure} QuantumCircuit
    \STATE \quad $circuit \; layout$: List[List[List[integer]]]
    \STATE \quad $unitary$: List[List[complex]]
    \STATE \quad $operations \; forward \; commuting$: List[boolean]
    \STATE \quad $gate \; counts$: List[integer]
\end{algorithmic}

\label{alg:sturcture-envstate}
\begin{algorithmic}[1]
    \STATE \textbf{Data Structure} EnvironmentState
    \STATE \quad $time \; step$: integer
    \STATE \quad $circuit$: QuantumCircuit
    \STATE \quad $action \; history$: List[integer]
    \STATE \quad $target \; circuit$: QuantumCircuit
    \STATE \quad $target \; action \; history$: List[integer]
    \STATE \quad $fidelity$: float
    \STATE \quad $rng \; key$: integer
\end{algorithmic}
\end{algorithm}

\begin{figure}[t] 
	\vskip 0.2in
	\begin{center}
		\centerline{\includegraphics[width=\columnwidth]{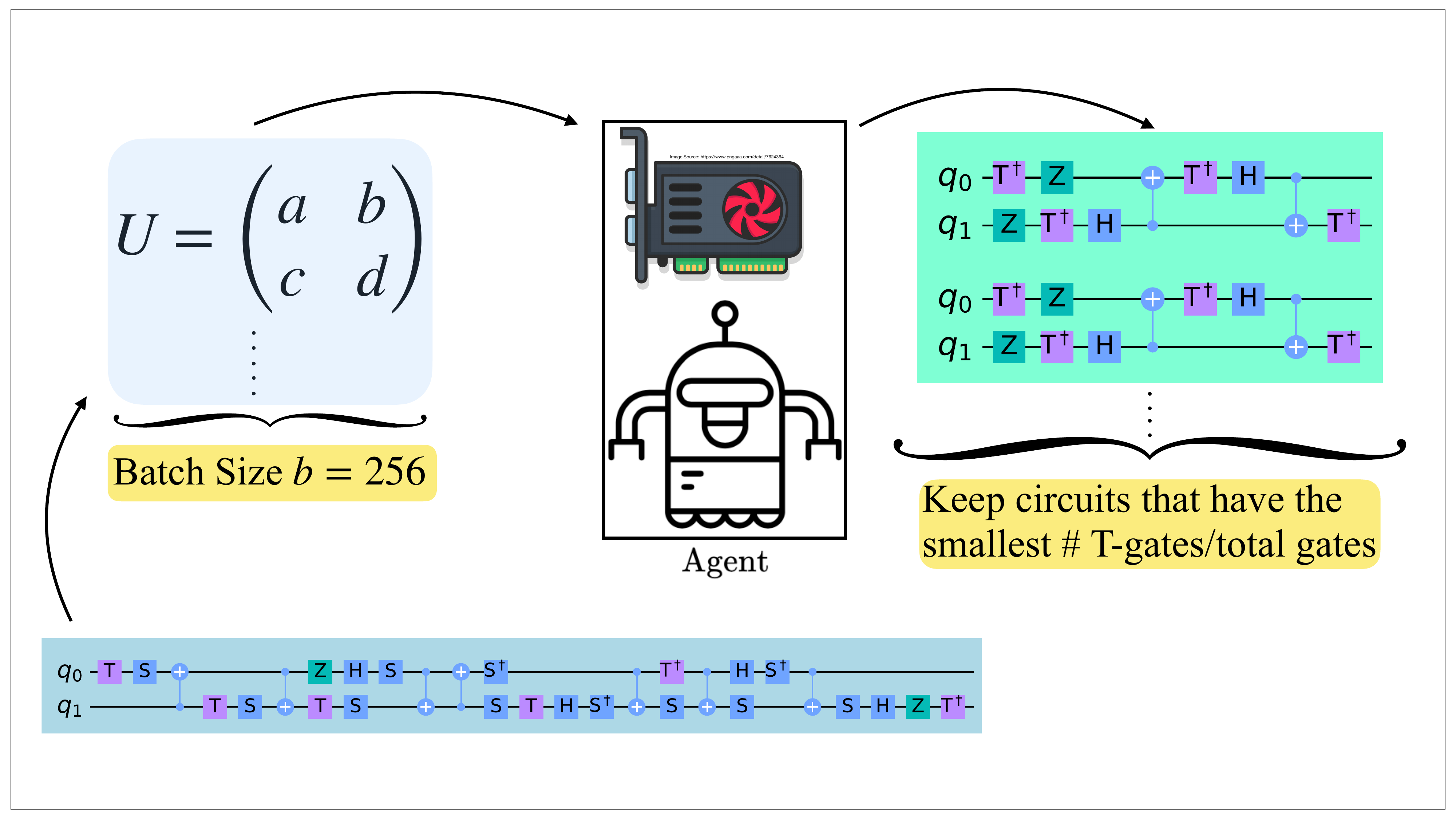}}
		\caption{Schematics of the Evaluation: The schematic representation of the evaluation of a single circuit performed by our \gls{rl} agent. We evaluate a given unitary in parallel and choose the resulting circuit with minimal \# T-gates/total gates.}
		\label{fig:evaluation_sch}
	\end{center}
	\vskip -0.2in
\end{figure}

\textit{c) Evaluation procedure: }
As noted in the previous paragraph, we leverage parallel search during evaluation. Here, a single target unitary is repeated 256 times, from which the best solution circuit is reported, in our case, circuits with the lowest gate count or the fewest T-gates (depending on the experiment). The schematic for evaluation is described in~\cref{fig:evaluation_sch}

\subsection{Results for Random Unitaries}
\label{sec: exps2}

For our evaluation of randomized unitaries, we sample 50 target unitaries from randomly generated circuits per data point. We do this using rejection sampling, where we sample random circuits until we find 50 distinct target unitaries with a specific number of T-gates or total gates. Specifically, we sample circuits with 0 to 20 T-gates and 3 to 60 total gates for 2, 3, and 4 qubits. For 5 qubits, we use the same range of T-gates but limit ourselves to 0 to 40 total gates. For comparison, we evaluate the same circuits using PyZX \cite{kissinger2019reducing}, an optimization heuristic based on ZX-calculus \cite{kissinger2020Pyzx}, which can reduce the number of T-gates in a circuit, and Synthetiq~\cite{paradis2024synthetiq} which is also a unitary synthesis tool. The results of PyZX optimization can be seen as heuristic upper bound on the T-count of a unitary if the corresponding circuit is known. We also compare the total number of gates for the three methods. The run-time of PyZX for the number of qubits we consider is less than a second, which is expected since it is an optimization heuristic. So, we do not compare its run-time with our RL agent.

\begin{figure*}[t]
  \centering
  \begin{tabular}{@{}c@{}}
    \label{fig:pyzx_RL_TCount}
    \includegraphics[width=0.9\textwidth,height=120pt]{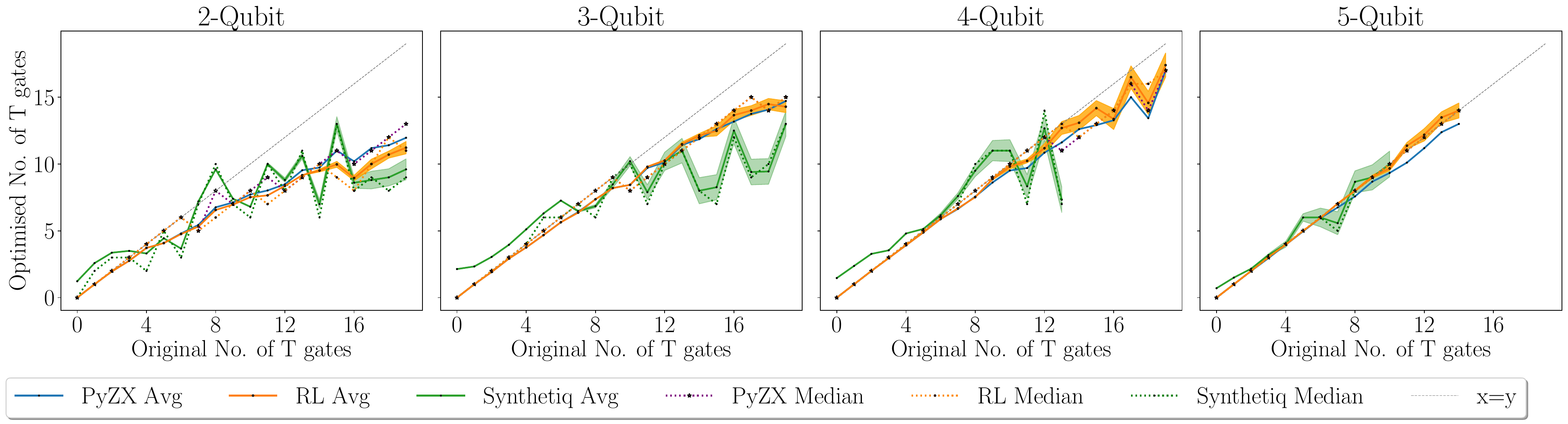} \\[\abovecaptionskip]
    \small (a) Comparison of average number of T-gates in the synthesised circuit by our RL agent, optimised circuit by PyZX and Synthetiq. 
  \end{tabular}

  \vspace{\floatsep}

  \begin{tabular}{@{}c@{}}
    \label{fig:circuits_distribution}
    \includegraphics[width=0.9\textwidth,height=100pt]{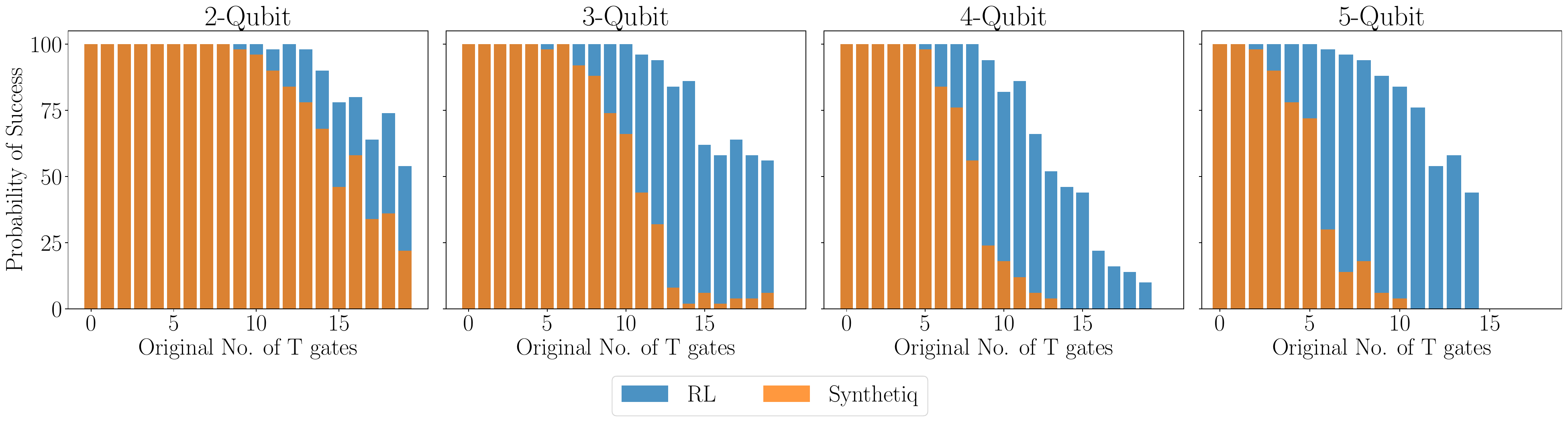} \\[\abovecaptionskip]
    \small (b) Distribution of unitaries our RL agent and Synthetiq successfully ($C_{HS} < 0.1$ and time-out = 400 s) synthesised.
  \end{tabular}

  \caption{Our RL agent's performance on synthesizing 50 unitaries, generated by randomly sampling Clifford+T gate set, building a circuit and calculating the corresponding unitary. Here, the number of T-gates for each unitary is fixed and ranges from 0 to 20 for 2, 3 and 4 qubits and 0 to 15 for 5 qubits. (a) Our agent can synthesize 2 and 3 qubit unitaries with similar average T-gate count as PyZX whereas it outputs slightly more number of T-gates for 4 and 5 qubits. Synthetiq on the other hand performs worse, and after around 10 T-gates the number of circuits synthesized drops significantly pulling down the average, giving a misleading impression of better T-gate count. (b) Here, we compare the successful synthesis of a given unitary with a given number of T-gates, as done by our RL agent and Synthetiq. We see a slight decline in performance for higher T-gate count for 2 and 3 qubit unitaries, and a significant drop for 4 and 5 qubit unitaries. The observed decline in performance can be attributed to the imbalance in the training set distribution. Specifically, there are fewer instances of circuits with larger T-gate counts compared to those with smaller counts.}\label{fig:Tcount}
\end{figure*}

\begin{figure*}[t]
  \centering
  \begin{tabular}{@{}c@{}}
    \label{fig:pyzx_RL_GateCount}
    \includegraphics[width=0.9\textwidth,height=120pt]{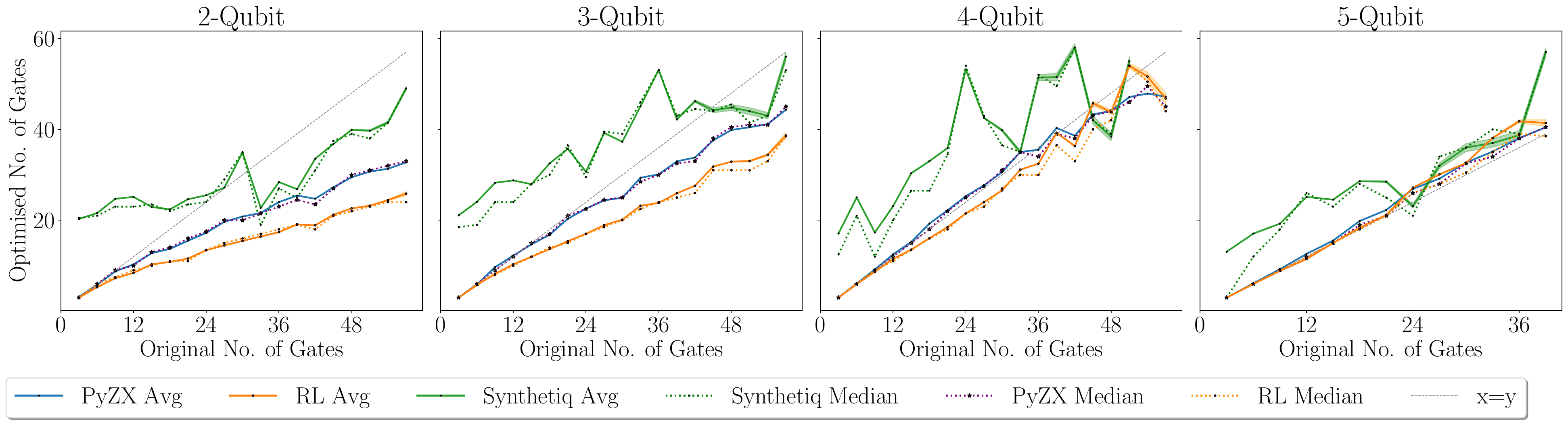} \\[\abovecaptionskip]
    \small (a) Comparison of the average total gate count in the synthesized circuit by our \gls{rl} agent, optimized circuit by PyZX and Synthetiq.
  \end{tabular}

  \vspace{\floatsep}

  \begin{tabular}{@{}c@{}}
    \label{fig:circuits_distribution_gatecount}
    \includegraphics[width=0.9\textwidth,height=100pt]{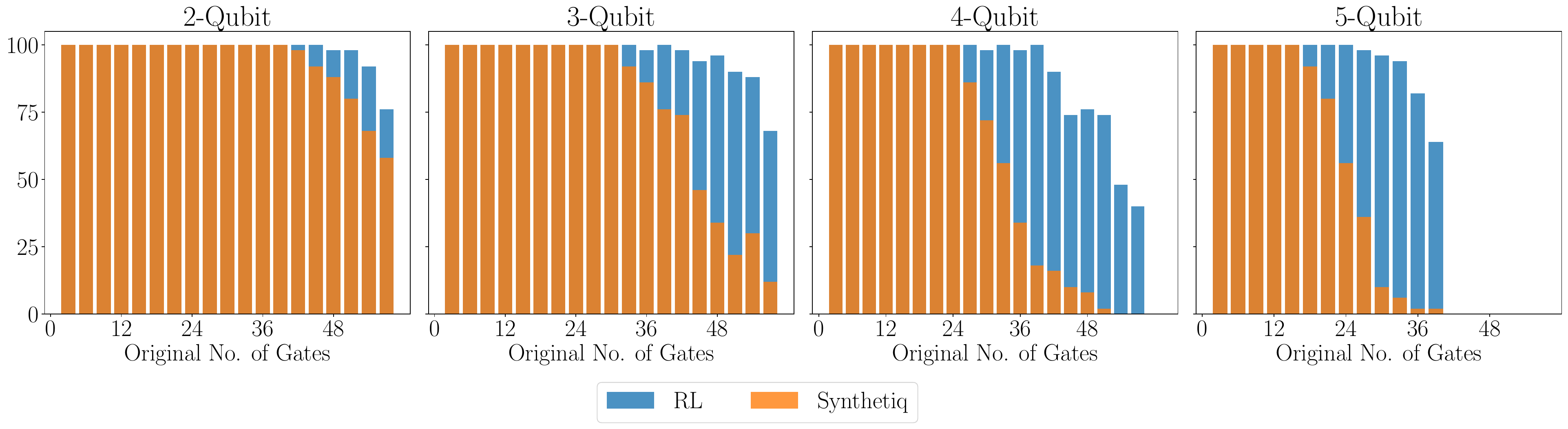} \\[\abovecaptionskip]
    \small (b) Distribution of unitaries our \gls{rl} agent and Synthetiq successfully ($C_{HS} < 0.1$ and time-out = 400 s) synthesised.
  \end{tabular}

  \caption{Our RL agent's performance on synthesizing 50 unitaries, generated by randomly sampling Clifford+T gate set, building a circuit and calculating the corresponding unitary. Here, the total number of gates for each unitary is fixed and ranges from 3 to 60 for 2, 3 and 4 qubits and 3 to 40 for 5 qubits. (a) Our agent can synthesize 2, 3 and 4 qubit unitaries with better average total gate count than PyZX and Synthetiq, although we see a drop in performance for higher total gate count compared to PyZX. Average total number of gates is similar for the case of 5 qubit unitaries. (b) Here, we compare the successful synthesis of a given unitary with a given number of T-gates, as done by our RL agent and Synthetiq. We see a slight decline in performance for 2 and 3 qubit unitaries, and a significant drop for 4 and 5 qubit unitaries for larger gate count for our RL agent, whereas Synthetiq performs worse than RL for larger number of gates.}\label{fig:gatecount}
\end{figure*}

In Fig.~\ref{fig:Tcount} we plot the performance of our RL agent in synthesizing random unitaries for a fixed number of T-gates. Fig.~\ref{fig:Tcount}(a) shows the average and median number of T-gates obtained by our RL agent over 50 random unitaries. Here, the line $x=y$ denotes the case where the number of T-gates in the input and the output circuits are the same. In Fig.~\ref{fig:Tcount}(b), we plot the fraction of unitaries successfully synthesized by our agent and Synthetiq. The success for our RL agent is determined by calculating $C_{HS}$ (\cref{eq:hilbert-schmidt}), for each of the 50 instances. When $C_{HS}>0.1$, evaluation is considered failed and $C_{HS}<0.1$ is considered success\footnote{Since PyZX is not a synthesis tool there is no possibility of failure for such small qubit count. PyZX is always able to either reduce the number of T-gates in the circuit or at least keep it the same.}. We empirically found the threshold of $0.1$ to work best for our experiments. Consequently, we verified whether the results with $C_{HS} < 0.1$ match the target unitary, which we found to be the case in all the experiments. Success rate for Synthetiq is determined by a time-out condition, in our case, 400 seconds on a 48 core processor. If no circuit is found within this time-limit the synthesis is deemed a failure. The error band shown in Fig.~\ref{fig:Tcount}(a) around the RL agent and Sythetiq curve is the fraction of unsuccessful runs out of the 50 unitaries considered. We see that the error band broadens for a higher number of T-gates, showing a decrease in performance.
In Fig.~\ref{fig:gatecount}, we show the performance for synthesizing unitaries for a fixed number of total gates.

We find that our \gls{rl} agent can synthesize all the target unitaries for small T-gate count and total gate count. The decrease in performance for T-gate count, can be attributed to the skewed distribution of T-gate numbers in the training set, meaning there are fewer circuits with larger T-gate counts given the current upper limit of 60 total gates. The performance decline for larger number of qubits can be attributed to the growing combinatorics for gate placement on the qubit lines along with higher dimensional observation vectors (number of elements of the unitary matrix). For 3, 4, and 5-qubit unitaries, our \gls{rl} agent performs similarly to PyZX in terms of the number of T-gates in the optimized circuit. The reduction in total gate count using our \gls{rl} agent was much better than PyZX since it has been observed~\cite{hietala2021verified} that PyZX T-gate optimization can increase the total number of gates. This highlights the benefits of our two-level optimization (both on the number of T-gates and total gates) procedure, which targets both the number of T-gates and the total gate count.
We also see the performance of Synthetiq against our RL agent on these randomly generated unitaries. The fraction of unitaries Synthetiq is able to synthesize degrades faster for deeper circuits compared to our RL agent both on the dataset of T-gates and total number of gates. For larger qubits and larger than 10 T-gates in the original circuit, the number of circuits Synthetiq is able to synthesise drops significantly as seen in Fig.~\ref{fig:Tcount}(b) which pulls down the average (Fig.~\ref{fig:Tcount}(a)), giving a misleading impression of better T-gate count in the synthesized circuit.

\subsection{Evaluation on Structured Unitaries}
\label{sec: exps1}

\begin{table*}[t]
    \caption{Benchmarking our method on structured unitaries against MIN-T-SYNTH \cite{mosca2021polynomial}, QCOpt ~\cite{nagarajan20212isiwqcsq} and Synthetiq~\cite{paradis2024synthetiq}. We compare the number of T-gates and the run-time of synthesis (`-' indicates algorithm was unsuccessful in finding the synthesis either due to time limit or no convergence to the solution). The example decompositions of some unitaries are given in \cite{amy2013meet}.}
    \centering
    \begin{tabular}{ll||ll||ll||ll||lll}
    \toprule
        \textbf{Qubits}          & \textbf{Unitary}                                                          & \multicolumn{2}{l||}{\textbf{Our RL agent}} & \multicolumn{2}{l||}{\textbf{MIN-T-SYNTH}} & \multicolumn{2}{l||}{\textbf{QCOpt}} & \multicolumn{2}{l}{\textbf{Synthetiq}} \\
        \midrule
                   &                                                                    & T            & Time(s)           & T            & Time(s)           & T        & Time(s)          & T        & Time(s)       \\

                   \midrule

                   \multirow{8}{*}{2} & CV                                                                 & 3            & 9.82             & -            & -                 & 3        & 4.16                 & 3            & 0.089        \\
                   & CY                                                                 & 0            & 9.50             & 0            & 0.224             & 0        & 16.84            & 0             & 0.084     \\
                   & CZ                                                                 & 0            & 9.41             & 0            & 0.330             & 0        & 0.1             & 0               & 0.08     \\
                   & SWAP                                                               & 0            & 9.27             & 0            & 0.353             & 0        & 0.01       & 0            & 0.08           \\
                   & W                                                                  & 2            & 9.62             & -            & -                 & 0        & 3221.21                       & 2             & 0.087        \\
                & CP                                                                 & 3            & 10.69             & 3            & 0.007             & 3        & 6.15                         & 3                 & 0.082\\
                   & CH                                                                 & 2            & 16.23             & -            & -                 & 2        & 147.12          & 2   
                    & 0.086                     \\
                   & CT                                                                 & -            & -                 & -            & -                 & -        & -                & -            & - \\

                    \midrule

        \multirow{8}{*}{3} & Toffoli                                                            & 7            & 25.92             & 7            & 4.99              & 7        & 10800               & 7               & 0.238                   \\
                   & \begin{tabular}[c]{@{}l@{}}Single\\ Negated\\ Toffoli\end{tabular} & 7            & 16.83             & 7            & 5.70              & -        & -                & 7        & 0.315              \\
                   & \begin{tabular}[c]{@{}l@{}}Double\\ Negated\\ Toffoli\end{tabular} & 7            & 16.78             & 7            & 3.39              & -        & -               & 7       & 0.252               \\
                   & Fredkin                                                            & 7            & 16.96             & 7            & 4.85              & -        & -        & 7          & 0.779     \\
                   & Peres                                                              & 7            & 16.87             & 7            & 5.00              & -        & -               & 7         & 0.321\\
                   & QOR                                                                & 7            & 16.56             & 7            & 5.09              & -        & -                & 7         & 0.250    \\
                   & AND                                                                & 7            & 16.83             & 7            & 3.44              & -        & -                & 7        & 0.498      \\
                   & TR                                                                 & 7            & 16.19             & 7            & 3.47              & -        & -        & 7        & 0.342              \\

                   \midrule
            \multirow{4}{*}{4} & \begin{tabular}[c]{@{}l@{}}3 Toffoli\\ (U2)\end{tabular}           & 7            & 18.54             & 7            & 309.84            & -        & -                & 7                   & $\boldsymbol{12.058}$              \\
                   & 1-bit adder                                                        & 7            & $\boldsymbol{18.53}$             & 7            & 312.55            & -        & -            & 8              &  24.365       \\

                   & 2 Peres           & 7             & 18.85              & 7            & 312.55             & -          & -               & 7             & $\boldsymbol{13.495}$                \\
                   
                   & \begin{tabular}[c]{@{}l@{}}2 Toffoli\\ (U1)\end{tabular}           & 12           & $\boldsymbol{19.27}$             & 11           & 6642.16           & -        & -            & -                 & -  \\
                   & \begin{tabular}[c]{@{}l@{}}Toffoli\\ (1 ancilla)\end{tabular}      & 7            & 31.50           & 7            & 233.54            & -        & -            & 7        & $\boldsymbol{4.132}$              \\
                \midrule
        \multirow{1}{*}{5} & x-Mod 5                & 12            & $\boldsymbol{60.773}$           & -            & -            & -        & -              & -        & - \\
                \bottomrule
    \end{tabular}
    \label{tab:SpecificUnitaries}
\end{table*}

For this experiment, we selected unitaries that are commonly encountered in various algorithms. These unitaries, such as Toffoli, Fredkin, controlled-Phase (CP), and controlled-$\sqrt{X}$ (CV) gates, are crucial for decomposing multi-qubit multi-controlled unitaries efficiently~\cite{barenco1995elementary}.

We compare our agent's performance with other synthesis algorithms and depict the results in~\cref{tab:SpecificUnitaries}. The table includes the number of qubits in the input circuit, the target unitary, the results obtained by our agent, and the results of MIN-T-SYNTH, QuantumCircuitOpt (QCOpt for short) and Synthetiq. For MIN-T-SYNTH, we used the code provided by the authors of Ref.~\cite{mosca2021polynomial}. While the evaluation was successful for some of our chosen examples, manual code adjustments would have been required, leading to `-' entries in the table. For QCOpt, the method exceeded the time limit of three hours without finding a solution in the basic Clifford+T gate set, which are also denoted by `-' in the table. For Synthetiq, we set a time limit of 800 seconds on a 48 core processor machine using all the cores, with `-' in the table if the time-limit is exceeded.

Our method successfully synthesized all the unitaries, except for the controlled-T gate, which was not synthesized by any of the algorithms. From Corollary 2 in Ref.~\cite{giles2013exact}, we know that controlled-T cannot be synthesized exactly without the use of an ancilla qubit. As per Refs.~\cite{amy2013meet, mosca2021polynomial}, our \gls{rl} agent achieves optimal T-count for all target unitaries except $U_1$, which was synthesized with one extra T-gate compared to MIN-T-SYNTH. We also see that Synthetiq out-performs our method in all the instances except for the 5-qubit x-Mod 5, 1-bit adder and $U_1$ in the time for synthesis. 

Both QCOpt and our \gls{rl} agent require an explicitly defined input gate set, which for $n$-qubit unitary would require $\mathcal{O}(n^2)$ possible ways to place CNOT gates and $\mathcal{O}(n)$ one-qubit gates unless we already have prior knowledge about the decomposition of the unitary. The difference between QCOpt and our method is that the mixed-integer programming (MIP) underlying QCOpt becomes highly inefficient in terms of run-time under such scaling. In comparison, our agent only has to account for these possibilities during training.

Comparing the run-times of our agent with the heuristic methods, we observed that, while MIN-T-SYNTH, QCOpt and Synthetiq are faster for simple unitaries, our agent takes approximately fifteen seconds for synthesis. However, as we analyze the increase in run-time with the number of qubits, we notice a significant increase in synthesis time for the other methods, whereas our method remains relatively constant. With Synthetiq, we see an increase in run-time by almost a factor of 10 with the number of qubits. Therefore, the most advantageous aspect of our RL algorithm is the inference time as we scale up till five qubits.

Our results are promising as they did not require handcrafted rules to handle the complexities of search-tree pruning and gate placements. Instead, we only leverage commutation rules and gate cancellation identities. \textit{Throughout training, our agent learns to account for these rules under the single incentive to reduce the total gate count.}

%% file: ApxQCOpt.tex
Here, we would like to clearly define the terminologies about Quantum compilation.
The umbrella term Quantum compilation accounts for the challenges faced by NISQ devices as well as Fault-tolerant devices, on 
both the software and the hardware level. The have been categorised as follows \cite{maronese2022quantum}:

\begin{enumerate}
	\item Translation of high level quantum algorithm into a hardware specific gate instruction set
	\item Qubit mapping and routing on connectivity constrained architecture
	\item Circuit optimization to reduce the depth of the circuit
	\item Scheduling of quantum operations on the real hardware
\end{enumerate}

We would like to focus here on the circuit optimization aspect of quantum compilation, whose primary aim can be described as
reducing the depth of the circuit by reducing the count of one or more specific gate type in the translated circuit from step 1 and 2. For example,
for the NISQ devices which use 2-qubit entangling gates and 1 qubit rotation gates, reducing the 2-qubit gate count is 
essential due to their high noise threshold \cite{masanes2002time}, whereas for fault-tolerant systems reducing the 
number of T-gates in the circuit takes priority \cite{eastin2009restrictions}. 
There are various approaches one can take in order to optimise a given quantum circuit. 

\begin{figure}
    \centering
    \begin{forest}
        for tree={font=\footnotesize, parent anchor=south, child anchor=north,  align=center}, before drawing tree={where n children=0{
                if={level()<4}{
                    tier=terminal,
                }{},
            }{},},
        [Circuit  optimization, s sep=0.05cm
            [Local \\ optimization, arr, for tree={fold}, before computing xy={s-=2mm}
                [Peephole \\ optimization]
                [Template \\ optimization]
                [ZX-calculus \\ based optimization]
                [AlphaTensor \\
                Quantum]]
            [Global \\ optimization, arr, l+=1mm, s sep=0.1cm
                [Unitary \\ Synthesis, arr, for tree={fold}, before computing xy={s-=2mm}
                    [MITM]
                    [MIN-T-SYNTH]]
                [Unitary \\ decomposition, arr, l+=1mm, before computing xy={s-=2mm}
                    [Cosine- \\ Sine]
                    [Quantum \\ Shannon]]]]
    \end{forest}
    \caption{Different levels of circuit optimization. Local optimization involves making modifications to a circuit on a local scale. For instance, ZX-calculus works with individual gates, while peephole optimization looks at a sub-circuit of the original circuit. On the other hand, global optimization takes into consideration the complete matrix representation of the original circuit. It aims to identify a more efficient circuit that can approximate or precisely match the original unitary.}\label{fig:tree}
\end{figure}
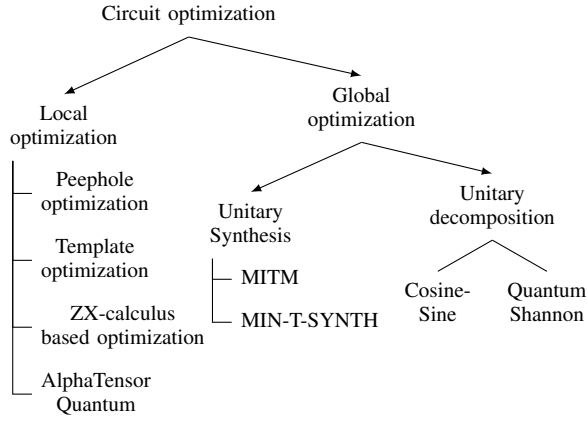

As shown in the tree diagram~\cref{fig:tree}, circuit optimization can be divided based on whether a local or a global optimization is performed. By local, we mean that the original quantum circuit is kept intact and modifications are done 
to a subset of qubits and small number of gates on those qubits, at a given time. Examples of such optimization algorithms 
are the peephole optimization, template optimization \cite{rahman2014algorithm, iten2022exact} and ZX-calculus based 
algorithm \cite{kissinger2020Pyzx}. A recent study using reinforcement learning for quantum circuit optimization was done in Ref.~\cite{ruiz2024quantum}, called AlphaTensor-Quantum, which can reduce the number of T-gates in a given circuit. A significant advantage of local optimization algorithms is that they can be applied 
for quantum circuits containing hundreds of qubits and gates without the use of extensive computational resources, but this 
also means that the optimality of the circuit is not guaranteed.

On the other hand, we have global optimization algorithms that take the entire high-level un-optimized quantum circuit in the 
form of its unitary matrix representation. Global optimization is further divided into unitary synthesis and decomposition. Although synthesis and decomposition have been used interchangeably in the literature, the difference becomes apparent if one looks at various decomposition and synthesis algorithms. One of the first papers on decomposing 
various unitaries and controlled unitaries~\cite{barenco1995elementary} provided prvable decompositions of the unitaries. Later, two-qubit decomposition~\cite{vidal2004universal, vatan2004optimal}, general three-qubit quantum 
gate decomposition~\cite{vatan2004realization}, cosine-sine decomposition~\cite{tucci1999rudimentary, mottonen2004quantum} 
and quantum Shannon decomposition~\cite{shende2005synthesis} for general multi-qubit gates and exact decomposition of 
multi-qubit Clifford+T circuits~\cite{giles2013exact} were provided. A common theme across all these 
approaches is that one can provably count the total number of unitaries required to build the target unitary. Alternatively, in other words,
the `decomposition' of the arbitrary target unitary has a known count in terms of the fundamental unitaries, which are well known.
Previous work on unitary synthesis, such as single-qubit unitary synthesis using Clifford+T~\cite{kliuchnikov2012fast}, 
depth-optimal synthesis using meet-in-the-middle (MITM)~\cite{amy2013meet}, matroid partitioning algorithm~\cite{amy2014polynomial},
COUNT-T algorithm~\cite{gosset2013algorithm,mosca2021polynomial}, QSEARCH, QFAST and LEAP~\cite{davis2020towards, younis2021qfast, smith2023leap}
where the primary aim is to use known basic gate sets (such as Clifford+T or CNOT and rotation gates) 
to approximate/represent the arbitrary target unitary, to reduce the total number of gates or the depth of the circuit, 
or a specific gate count (number of T-gates or CNOT gates) without getting the exact 
number of gates in a general sense.
This distinction between unitary synthesis and decomposition is crucial to make things straightforward for 
future work.